# Development of Photonic Crystal Fiber Based Gas/ Chemical Sensors

**Ahmmed A. Rifat[1,*], Kawsar Ahmed[2], Sayed Asaduzzaman[3], Bikash Kumar Paul[3], and Rajib Ahmed[4]**

[1]Nonlinear Physics Centre, Research School of Physics & Engineering, Australian National University, Acton, ACT, 2601, Australia

[2]Department of Information and Communication Technology, Mawlana Bhashani Science and Technology University, Santosh, Tangail-1902, Bangladesh

[3]Department of Software Engineering, Daffodil International University, Sukrabad, Dhaka-1207, Bangladesh

[4]Nanotechnology Laboratory, School of Engineering, University of Birmingham, Birmingham B15 2TT, UK

*E-mail: RifatAhmmed.Aoni@anu.edu.au

**Abstract:** The development of highly-sensitive and miniaturized sensors that capable of real-time analytes detection is highly desirable. Nowadays, toxic or colorless gas detection, air pollution monitoring, harmful chemical, pressure, strain, humidity, and temperature sensors based on photonic crystal fiber (PCF) are increasing rapidly due to its compact structure, fast response and efficient light controlling capabilities. The propagating light through the PCF can be controlled by varying the structural parameters and core-cladding materials, as a result, evanescent field can be enhanced significantly which is the main component of the PCF based gas/chemical sensors. The aim of this chapter is to (1) describe the principle operation of PCF based gas/ chemical sensors, (2) discuss the important PCF properties for optical sensors, (3) extensively discuss the different types of microstructured optical fiber based gas/ chemical sensors, (4) study the effects of different core-cladding shapes, and fiber background materials on sensing performance, and (5) highlight the main challenges of PCF based gas/ chemical sensors and possible solutions.

## 12.1 Introduction

Optical sensor devices have been taken as an alternative to conventional solid-state planar, brittle, less flexible, and rigid electronic devices [1]. Electronic devices have some major limitations such as high manufacturing cost, complex procedure, slower response time and reliability as compared to the optical sensors. Electronic devices are also affected with electromagnetic (EM) and thermal noise or interference [2]. Now a days, physical sensing based on optical platform used to sense and monitor complex environment and its surrounding such as temperature, humidity, strain, stress, pressure, and torsion, etc. having important applications in wearable sensors, robotics, health and safety monitoring [3-8]. Therefore optical sensor devices have been found the suitable alternative for the gas, chemical and oil sensing applications, due to its advantages of low cost, less noise/interference, higher sensitivity, fast response, reliability, and compactness [9-11].

Since last decades, photonic crystal fiber has been shown great development in optical sensing [12-15]. Due to advance optical instrumentations, the field of fiber optics is no longer limited to telecommunication applications. PCF also known as holey fiber, consists of periodically ordered microscopic cylindrical air holes running through the full length of the fiber. The standard PCF is made with fused-silica ($SiO_2$) that has a regular pattern of voids or air holes that run parallel to its axis. Unlike traditional optical fibers, both the core and cladding are made from the same material. The structural view of PCF shown in Fig. 12.1. Due to unique advantages of PCFs such as design freedom, light controlling capabilities, faster detection response and miniaturized structure; it has been received considerable attention in developing opto-devices and sensors. Moreover, modifying the structural parameters of PCFs, such as air-holes diameter, pitch size and number of rings, evanescent field can be controlled as a results it will find the large scale of possible applications, especially in sensing.

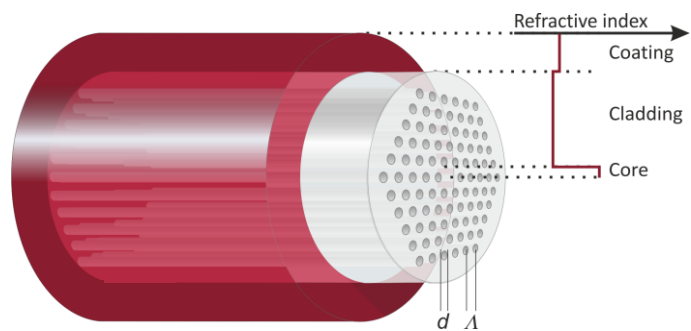

**Fig. 12.1** Standard Photonic Crystal Fiber Structure.

The first working PCF emerged from the drawing tower in late 1995. PCF was practically fabricated for the first time in 1996 by Philip Russel [16]. The first fabricated PCF is shown in Fig. 12.2a. There are mainly two types of PCF, one is Index Guiding PCF and another is Photonic Band Gap (PBG) PCF. The cross-section diagram of index guided photonic crystal fiber shown in Fig. 12.2b. Photonic-bandgap guiding occurs by surrounding the core of an optical fiber with the cladding region which contains air holes running along the length of the fiber. The cross sectional diagram of photonic-bandgap is shown in Fig. 12.2c. Recently, researchers are more interested in special type of PCF where both core and cladding are micro-structured (see Fig. 12.2d).

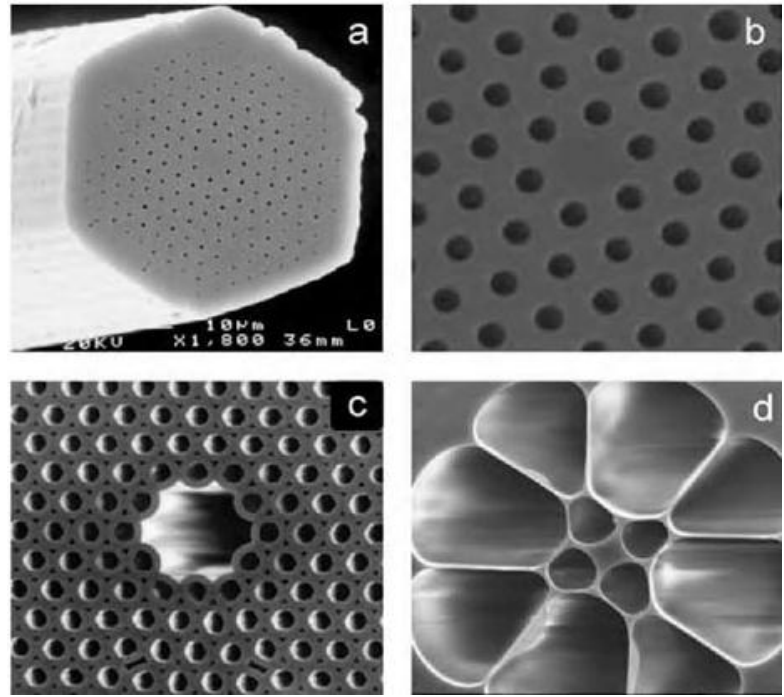

**Fig. 12.2** (a) The first working PCF. (b) Low loss solid core PCF. (c) The first hollow-core PCF. (d) A small core PCF extruded from Schott SF6 glass [16, 17].

Nowadays, PCF has been attracted much attention for its incredible performance and broad range of applications such as filters [18], switches [19, 20], electro-optical modulators [21, 22], polarization converters [23], and sensors [24-28], etc. Since last few decades, PCF has been considered and widely investigated as a suitable candidate for the optical sensing. Highly sensitive liquid and gas sensors are playing an important role in industrial processes especially for detecting toxic and flammable gasses or liquids to overcome the safety issues [29]. So, it has become one of the key challenges to enhance the performance of liquid and gas sensors. Photonic crystal fiber based liquid and gas sensors have been shown excellent performance in terms of sensitivity response. In recent years, researchers have been shown great interest on the development of PCF based sensors for environmental and safety monitoring [30, 31]. A wide variety of PCF based sensing techniques have been reported by changing different geometric parameters of the PCF to get higher sensitivity, detection accuracy and faster response time. Performance of the PCFs can be enhanced by regular or irregular geometric structure like hexagonal [32], octagonal [33], decagonal [34], square [35], honeycomb cladding [36], elliptical [37], and kagome [38]. Development of regular or irregular PCF structure leads to achieve more efficiency as well as use it in multipurpose like gas sensing, chemical sensing, bio sensing, cancer cell detection, medical science, temperature sensing, illuminations, machining, and welding applications because of its smaller size, lighter weight, chemically inertness, higher bandwidth, longer repeater span, electromagnetic immunity, and many other intriguing properties [39-41].

Researchers are trying to improve the performance of the PCF based sensors following different geometric shape of cladding and filling different transparent material in the core and cladding. However, doping material will cause high cost and also led to the complex fabrication process. As a result, considering practical point of view, a simple PCF structure is required with high sensitivity virtues. Park *et al.* [42] proposed a new type of index-guided PCF to enhance sensing capability by introducing a hollow high index ring defect that consists of the central air-hole surrounded by a high index $GeO_2$ doped $SiO_2$ glass ring. They showed that fraction of evanescent field was increased by increasing the diameter of central air hole, hence the sensitivity was improved and kept the diameter smaller than cladding holes diameters for the physical realization of effective index guiding. Olyaee *et al.* [43] showed sensitivity 13.23% by increasing the diameter of the holes located in the inner ring and reduced confinement losses to $3.77 \times 10^{-6}$ dB/m by increasing the air-holes diameters located in the outer ring at the wavelength 1.33 μm. Ademgil *et al.* [44] proposed a microstructured-core and cladding PCF for liquid sensing and found the sensitivity of 20.10% at the wavelength λ=1.33 μm. According to the reported papers, it has been observed that both relative sensitivity and confinement loss are improved. But these are not the significant value for a gas sensor. To increase the relative sensitivity and lower the confinement loss more research has to be done. An effective way to increase the performance of gas sensor is to design a simple PCF structure which will allow more penetration of the evanescent fields.

In this chapter, sensitivity and guiding properties of various index-guiding PCFs based gas/ chemical sensors as well as liquid sensors are extensively discussed. The finite element method (FEM) with perfectly matched boundary layer (PML) conditions are extensively used for the computational study of PCF based gas/ chemical sensors, as a result, we also described the PML effect on sensing. The important sensor parameters such as sensitivity and confinement loss effect due to change of pitch, number of rings, air-hole diameters, and air-filling fractions (AFF) are investigated. Recent advances of existing PCF based gas/chemical sensors are discussed, which consist of comparisons among several PCF structures in terms of relative sensitivity, and fabrication feasibility. Finally, research gaps of this field are addressed and potential future detections to overcome them are discussed.

## 12.2 Fundamentals of PCF Based Sensors

### 12.2.1 Sensing Mechanism of PCF Based Sensors

The criteria's for which PCFs are used as a sensor-

- ➢ **Absorbance**, measured in a transparent medium.
- ➢ **Reflectance**, measured in non-transparent media, usually using an immobilized indicator.
- ➢ **Luminescence**, based on the measurement of the intensity of light emitted by a chemical reaction in the receptor system.
- ➢ **Fluorescence**, measured as the positive emission effect caused by irradiation.
- ➢ **Refractive index**, measured as the result of a change in solution composition.
- ➢ **Opto-thermal effect**, based on a measurement of the thermal effect caused by light absorption.
- ➢ **Light scattering**, based on effects caused by particles of definite size present in the sample.

Among these criteria's, absorbance property is widely used for sensing applications which followed by the absorption spectroscopy for practical realization. In recent years, absorption techniques are also used to detect the gas/ chemicals. We have discussed about the absorption spectroscopy in the following paragraph.

**Fundamentals of Absorption Spectroscopy**

The quality of gas assimilation lines can be utilized to perform quantitative estimation of gas. The sensitivity of the analytes is derived from the output optical power and it is followed by the Beer-Lambert law [43, 45] as follows.

$$I(\lambda) = I_0(\lambda) \exp(-r\alpha_m l\, c) \qquad (12.1)$$

where $I_0$ is the emerging optical intensity of light passing through the targeted analyte and $I$ is the intensity of light for investigation. The relative sensitivity coefficient is r and $\alpha_m$ is the absorption coefficient and $l$ is proportional to the line strength, and $c$ is the concentration of the sample analyte.

The absorbance of the sample is defined as [25];

$$A = -log_{10}\left(\frac{I_0}{I}\right) = r\alpha_m lc \qquad (12.2)$$

The molecule shows different properties in their different states. Some gases are extremly flammable and hazardous that can be detected by PCF. The flammable gasses may cause explotion and fire in any industries as well as residential areas. Some gases are toxic that may cause differecnt types of diseases. Few gases may cause cancer and other disorders as well. The table 12.1 shows the detailed description and absorption wavelength.

**Table 12.1** Absorption wavelength and line strength of some common gases [45]

| Molecule | Absorption wavelength (μm) | Line strength ($cm^{-2}atm^{-1}$) | Descriptions |
|---|---|---|---|
| Acetylene ($C_2H_2$) | 1.533 | ~20×$10^{-2}$ | Extremely flammable |
| Hydrogen iodide (HI) | 1.541 | 0.775×$10^{-2}$ | Highly toxic, colorless |
| Ammonia ($NH_3$) | 1.544 | 0.925×$10^{-2}$ | Toxic, irritating and destructive to tissues |
| Carbon monoxide (CO) | 1.567 | 0.0575×$10^{-2}$ | Combustion product, toxic, colorless |
| Carbon dioxide ($CO_2$) | 1.573 | 0.048×$10^{-2}$ | Main greenhouse gas |
| Hydrogen sulfide ($H_2S$) | 1.578 | 0.325×$10^{-2}$ | Toxic, colorless, flammable |
| Methane ($CH_4$) | 1.667, 1.33 | 1.5×$10^{-2}$ | Flammable, greenhouse gas |
| Hydrogen fluoride (HF) | 1.330 | 32.5×$10^{-2}$ | Toxic, colorless, extremely corrosive |
| Hydrogen bromide (HBr) | 1.341 | 0.0525×$10^{-2}$ | Highly toxic, colorless |
| Nitrogen dioxide ($NO_2$) | 0.800 | 0.125×$10^{-2}$ | Greenhouse gas |
| Oxygen ($O_2$) | 0.761 | 0.01911×$10^{-2}$ | Strong oxidizer, supports and vigorously accelerate combustion |

The PCF sensing mechanism depends on the absroption lines of the corresponding gases. By the effective refractive index of the related gas which can be detected shows modal intensities at the core region. Gas species with absorption in near IR region and line strength are listed in Table 12.1. By the absorption cell, gases can be detected between any ranges of wavelength (see Fig. 12.3).

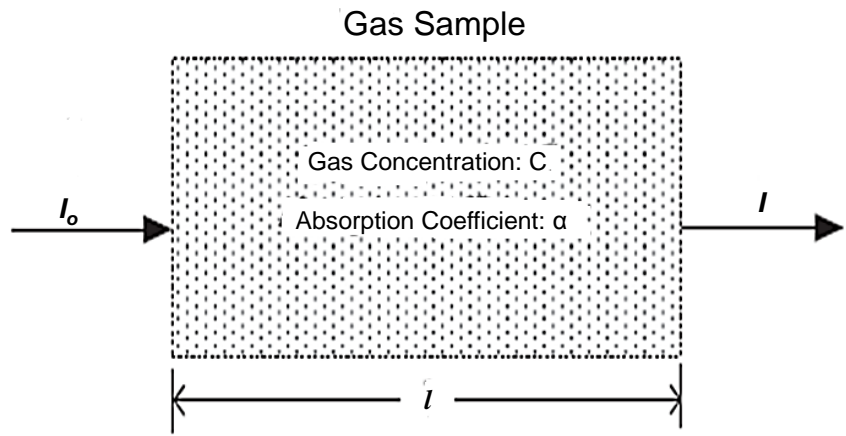

**Fig. 12.3** Guided Absorption Cell.

Fig. 12.4 exhibits the schematic block diagram of the checked wavelength direct absorption spectroscopy system, which is used for methane ID. Light from a tenable laser source (TLS) is dispatched into a Solitary Mode Fiber (SMF). The SMF is butt-coupled to the HC-PBF using a 3-turn positioner. An opening is left between the terminations of the fibers to allow the gas access into the focal point of the HC-PBF. The other side of the HC-PBF is joined to a SMF plait using a business roundabout section splicer as described in [45]. The incapacitating of this splicer is around 1 dB. The light transmitted through the HC-PBF is measured using a Ge-identifier. A PC is used to control the tuneable laser and accumulate data from power meter which contain the locator. The open end of the HC-PBF is set in a settled vacuum chamber. Finally, a pump is used to clear the fiber before stacking with the adjusted joining of the targeted gas.

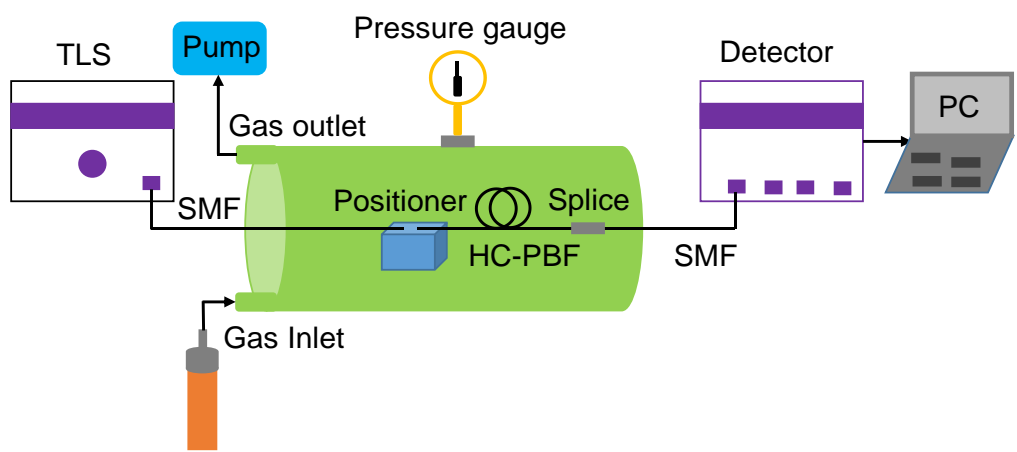

**Fig. 12.4** Direct absorption spectroscopy system [46].

To measure the methane maintenance territory use the two HC-PBF cells. The setup of absorption spectroscopy showed up in the above figure (Fig. 12.4). Consequently, the vacuum chamber is required to clear first then it is stacked with a balanced mix of 18,750 ppmv (parts per million volumes) methane in air at a relative weight of 1 bar.

Resulting to allow sufficient time for the gas to thoroughly diffuse into the fiber, a transmission extent is recorded. The intentional territory is then established to a reach accumulated with void cells, yielding the ingestion in light of methane. From the methane maintenance band, it can be seen that the methane has generally more grounded absorption at 1331.55 nm.

### *12.2.2 Applications of PCF Based Sensors*

Due to unique optical properties of PCFs, it has been found large scale of potential applications. There are many applications in PCF based sensors and some of them are listed below:

- **Gas sensor**: Gases are colourless and can be toxic. But different gas has different absorption line based on its absorption spectrum length as well as refractive index. PCF has been used to detect those colourless gases [47]. PCF based gas sensor has the capability to detect CO that is commonly known as silent killer [43].
- **Chemical Sensor**: Chemicals are massively used in the industrial applications. In some cases it is bounded to detect some unwanted chemical those are poisonous for human body. Based on the internal structure of the chemical their refractive index are also different. Each chemical has its unique refractive index like for Benzene (n=1.366), Ethanol (n=1.354), water (n=1.33). Based on these refractive index chemical can be detected by PCF by passing the chemical through the core region [48].
- **Biosensor**: Biosensor is a device that can senses the numerous biological molecules and/ or antibodies, enzymes with presence of associated chemical or analyte. Nowadays, PCFs are used to detect biological substances like urine glucose, $P^H$, serum protein, etc. [49].
- **Temperature Sensor**: Temperature measurement is a key issue in industrial and environmental health monitoring purposes. The conventional temperature measurement mechanism is not suitable. PCF based temperature sensors are most popular for temperature sensing because of its simple and cost effective detection capabilities. To date, several number of PCF based temperature sensors have been reported [50, 51].
- **Refractive Index Sensor**: Refractive index based sensors are vastly using in optical sensing area. Based on the refractive index of different specimen, different applications are already employed among them Surface Plasmon Resonance (SPR) is notable [52, 53].

- ❖ **Corrosion Sensor**: The deterioration and loss of a material is commonly known as corrosion. With the development of PCF based corrosion sensors it can be easily monitored the present structural condition of aircraft [54], steel [55] and other materials.
- ❖ **Pressure Sensor**: Pressure is a common behaviour in cases of gaseous and liquid chemical. Moreover, the revolutionary changes in the optical field, PCF based pressure sensor played the significant role by measuring the pressure inside the patient body especially for urodynamic and cardiovascular assessment [56].
- ❖ **Humidity and Moister Sensor**: To predict the present state of the atmospheric condition some weather forecasting parameters are needed to be calculated. Among them humidity is exceedingly crucial. Nowadays, PCF based humidity and pressure sensors are using widely [57, 58].
- ❖ **Flying Particle Sensor**: Different particle are generally presence in the environment (like in open air). But the particles are too tiny that it cannot be easily detected. Some radioactive substance may be presence in the different areas air and those are flying randomly. This type of sensor can be used in radioactive areas to monitor the environment [59].
- ❖ **Transverse Load Sensing**: Utilizing the mechanism of light propagation property load can be sensed. Optical fiber can directly use to sense load in the transverse direction [60].

### 12.2.3  *Advantages of PCF Based Sensors*

By varying PCF structure (like hexagonal, octagonal, decagonal, etc) and also varying the structural core-cladding size with different arrangement, propagating light can be tailored significantly. For the PCF based sensors, evanescent field is the key element. Due to PCFs light controlling capability, evanescent field can be tailored intensely, as a result sensor performance can be improved significantly.

### 12.2.4  *Optical/ Guiding Properties of PCF Sensors*

The better optical guiding properties ensure the application of PCFs as a gas/chemical sensors. Optical properties are the core area of research on PCF. The sensor guiding properties of the PCFs are discussed below step wise in details [12, 61].

**Relative Sensitivity**

First and foremost nature of a PCF used as sensor is needed to compute the relative sensitivity response. Relative sensitivity response of a PCF denoted that the sensing capacity of the proposed PCF. Relative sensitivity is symbolized by r and it can be calculated through the following equation [47];

$$r = \frac{n_s}{\text{Re}[n_{eff}]} f \qquad (12.3)$$

where $n_s$ is the refractive index of target gas species, typically consider as 1 and Re[$n_{eff}$] is the real part of the effective mode index. Here, f is the fraction of holes power by total optical power which can be defined as [62]:

$$f = \frac{\int_{holes} \text{Re}(E_x H_y - E_y H_x) dx dy}{\int_{total} \text{Re}(E_x H_y - E_y H_x) dx dy} \qquad (12.4)$$

Here, $E_x$, $E_y$, $H_x$ and $H_y$ are the transverse electric and magnetic field of the guided mode, respectively.

**Confinement loss**

Confinement loss or leakage loss occurs due to leaky nature of the mode and irregular arrangement of air holes. Those air holes are playing the role of dielectric medium. Confinement loss also depends on transmitted wavelength, parameter shape and size, number of holes, and rings. A circular shaped anisotropic perfectly matched layer (C-APML) is use to satisfy the boundary condition which avoids unwanted electromagnetic reflection at the boundary of PCF. By this term, confinement loss or leakage loss can be calculated by the imaginary part of the effective refractive index. The confinement loss or leakage loss can be calculated by the following equation [63]:

$$L_c \, [dB/m] = 8.686 K_0 Im(n_{eff}) \times 10^6 \qquad (12.5)$$

Where, $K_0 = 2\pi/\lambda$, is the wave number and Im[$n_{eff}$] is the imaginary part of the effective refractive index.

**Birefringence**

Birefringence is one of the crucial properties of PCFs. It is highly influential for polarization maintaining fiber (PMF). Birefringence is a property of a PCF comes from some geometric asymmetry based on air holes position. Highly structural asymmetry of PCF, especially 1st ring of the PCF produces higher order of birefringence and structural symmetry of PCF has no influence to produce birefringence. The mathematical formulation of birefringence can be expressed as [64],

$$B(\lambda) = |n_{eff}^x - n_{eff}^y| \qquad (12.6)$$

**Beat length**

Another wavelength dependent parameter is beat length. Beat-length is a significant argument to discover the birefringent optical fibres. It defines the optical signal transmission length along the fiber when the phase difference of two orthogonal polarization states varies 360 degrees or $2\pi$ radians. This property leads to periodic power exchange between two orthogonal components. This period is called beat length which can be evaluated by the following expression [65]:

$$L_B(\lambda) = \frac{\lambda}{B(\lambda)} \qquad (12.7)$$

**V-Parameter**

A PCF can be single mode or multimode. A cut-off value always present there for determining the fiber for modal analysis. If the value of V is less than or equal to 2.405, then it indicates the single mode operations, otherwise, it permits multimode operations. The single mode response for step index fiber can be determined by the V-parameter which is defined by [66]:

$$V_{eff} = \frac{2\pi}{\lambda} a \sqrt{n_{co}^2 - n_{cl}^2} \qquad (12.8)$$

Here, $n_{co}$ and $n_{cl}$ are the refractive index of core and cladding; a is the radius of the fiber core. The fraction of optical power in a certain mode is bounded inside a fiber core determined by V number. The lower V-value indicates the optical power fraction is low and vice-versa.

**Effective Area**

Effective mode area is generally considered as the light carrying region. For fundamental propagating mode, electric-field (E) distribution occurs inside the core, as a result, effective mode area (EMA) of a PCF can be determined by the following equation [32]:

$$A_{eff} = \frac{(\iint |E(x,y)|^2 dxdy)^2}{\iint |E(x,y)|^4 dxdy} \qquad (12.9)$$

For high bit rate data transmission system (especially for telecommunication), large effective mode area is required. On the contrary, lower EMA is preferable for nonlinear applications.

**Nonlinearity**

High optical power density is provided by a small effective area for which the nonlinear effects would be significant. The nonlinear effective or nonlinearity is closely related with the effective area and also nonlinear coefficient of the PCF background material in associated with the operating wavelength λ. The nonlinear coefficient can be examined by the following equation [32]:

$$\gamma = \left(\frac{2\pi}{\lambda}\right)\left(\frac{n_2}{A_{eff}}\right) \qquad (12.10)$$

where $n_2$ is the nonlinear refractive index. Nonlinear effects are very advantageous in different optical devices and optical applications such as broadband amplification, channel demultiplexing, wavelength conversion, soliton formation, optical switching and many more applications. Nevertheless higher orders of nonlinearity are responsible for supercontinuum generation (SCG).

**Splice Loss**

Splice loss is another important parameter for fiber design consideration. Generally, for longer distance signal carrying or longer distance optical communication aspect two fibres are experienced by joining or splicing. It is very sensitive issue because due to small mismatch of the fibres during the splicing will led to the large signal attenuation. Splice loss occurs during the splicing between PCF and the single mode fiber. Splice loss can be calculated by the following equation [32]:

$$L_S = -20 log_{10} \frac{2 W_{SMF} W_{PCF}}{W_{SMF}^2 + W_{PCF}^2} \qquad (12.11)$$

where $W_{SMF}$ and $W_{PCF}$ are the mode field diameters of the single mode fiber and PCF, respectively.

**Refractive Index**

Transparent materials are highly used for PCF fabrication. The common transparent material silica has been used extensively for the PCF fabrication. Even as a standard PCF, silica fiber has been considered till now. Due to the technological advancement, different types of transparent materials such as Tellurite, Graphene, ZABLAN and TOPAS have also been shown great interest. These all materials basic characteristic are completely depends on refractive index. Refractive index is a material itself property which can be increased or decreased by doping other materials based on the different applications.

**12.3  Overview of PCF Based Gas/ Chemical Sensors**

### 12.3.1 Conventional Optical Fiber Sensors

Optical fiber based surface plasmon resonance (SPR) sensor has been reported by Mishra et al. [67], for the detection of hydrogen sulphide gas. The schematic diagram of the experimental set-up shown in Fig. 12.5. To utilize the SPR based gas sensor, nickel oxide doped ITO thin has been used. Gas chamber having the facility with inlet and outlet was used and the fiber probe was inserted the gas chamber. The un-polarized light lunched one end of the fiber and other end of the fiber connected to the spectrometer. To study the SPR response of the hydrogen sulphide gas, the gas chamber was evacuated with the help of a rotary pump and the reference signal was recorded.

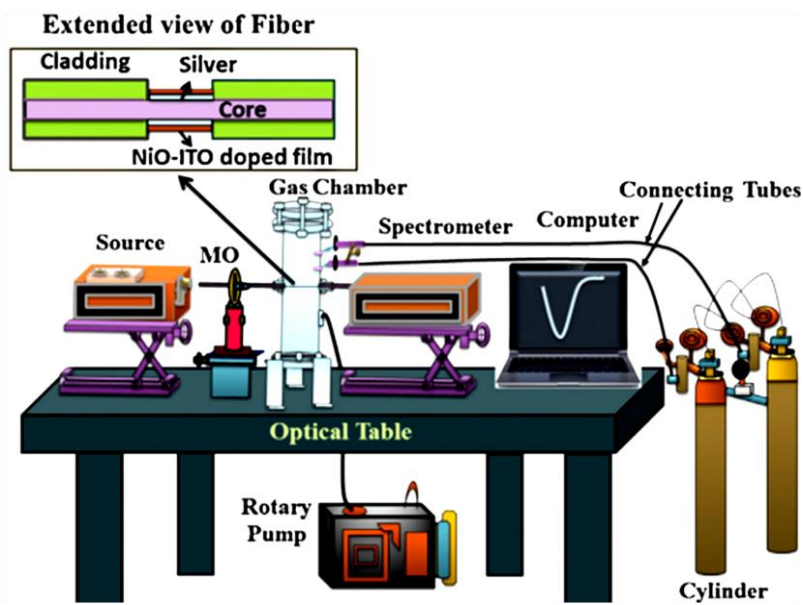

**Fig. 12.5** Conventional optical fiber sensor [67].

Recently, optical fiber based SPR sensor for the detection of Ammonia gas has also been reported [68]. The sensitivity of the sensor with optimized thickness of BCP layer is 1.891 nm/ppm and is larger than the sensitivity values obtained in the cases of Ag/BCP and Cu/BCP-coated probes for the concentration range 1–10 ppm of the ammonia gas

### 12.3.2 PCF Based Sensors

Based on geometrical structure, different types of PCFs have been reported for the gas/ chemical sensing. The most common and modest PCF based sensor structure is hexagonal structure. Moreover, circular, square, octagonal structures have also been reported. These types of shapes are constructed based on the orientation of air-holes in the cladding region. For the various architectural shapes light guiding mechanism through the fiber may have different manner. Here, from the design prospective, different types of PCF based sensor are shown in Fig. 12.6. Fig. 12.6 shows the different types of PCF such as octagonal (Fig.12.6a), hexagonal (Fig.12.6b), hybrid (Fig.12.6c), spiral (Fig.12.6d), hybrid-Combined (Fig. 12.6e) and square (Fig. 12.6f) based chemical/ gas sensors.

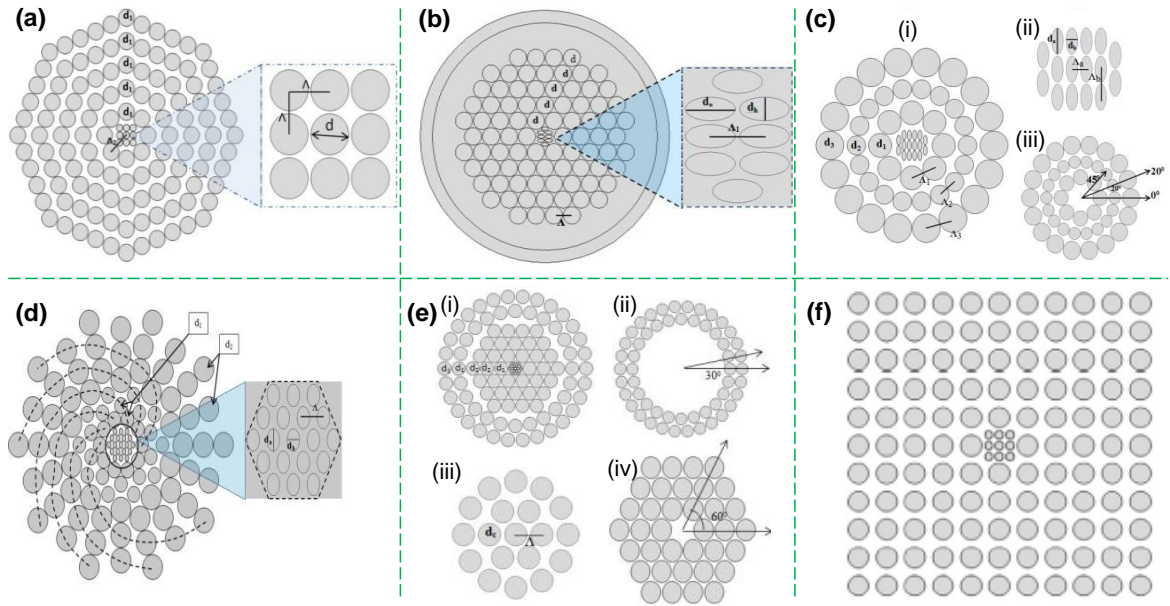

**Fig. 12.6** Different types of PCF based Sensors. (a) Octagonal, (b) Hexagonal, (c) Hybrid, (d) Spiral, (e) Hybrid-Combined, and (f) Square.

## 12.4   Guiding Properties Controlling Parameters of PCFs

Guiding properties or so called optical properties can be controlled by some fundamental parameters of PCF. By recent vibrating research it has been concluded that pitch, air filling ratio, diameters of air holes of both core and cladding significantly affect the result of wave guiding properties. In this segment a brief discussion about the controlling fundamental parameters and their effect.

### *12.4.1 Pitch Effects on Sensing*

In PCF, air holes exists through the entire fiber those are forming on silica background. These air holes are organized in a well-defined geometrical pattern. The hole to hole distance more specifically centre to centre distance of two adjacent air holes is called pitch (see Fig. 12.7). By altering pitch sensitivity of PCFs can improve.

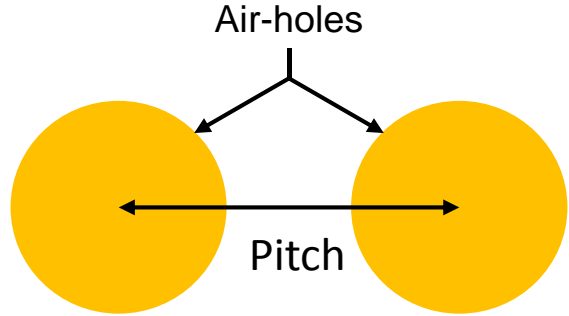

**Fig. 12.7** Pitch of PCF.

Fig. 12.8 shows the pitch effect on relative sensitivity of PCF. The figures (see Fig. 12.8a, 8b & 8c) illustrates that smaller the pitch led to the higher relative sensitivity. Because the lower pitch value induces the lower space between air holes which result congested air holes. These air holes direct the evanescent field through the core region.

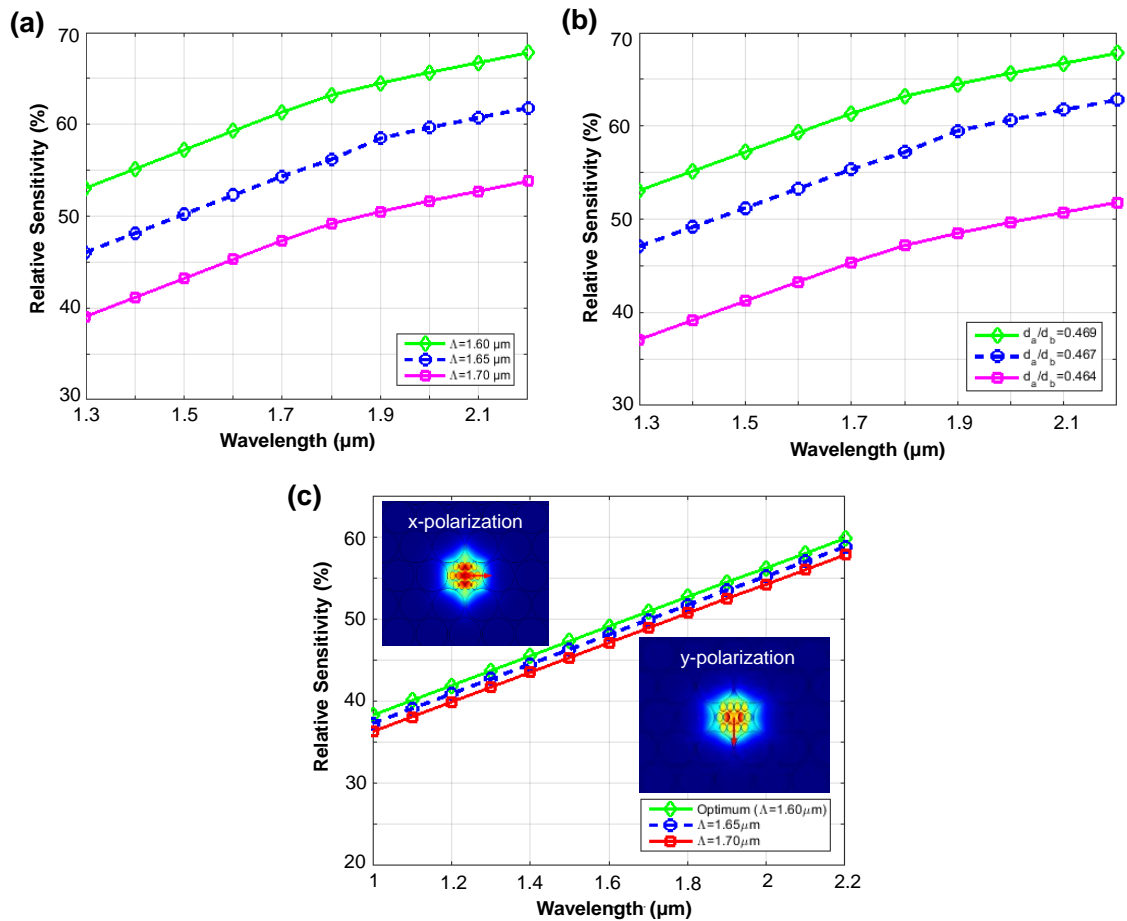

**Fig. 12.8** Analysis the pitch effect on relative sensitivity of PCF [12].

### 12.4.2    Diameter Effects on Sensing

Diameter has significant effects on relative sensitivity. These diameters means air holes diameter of core region as well as cladding region. By changing the diameters relative sensitivity can tailored. In Fig 12.9a & 12.9b, it is observed that the larger diameters shows higher relative sensitivity.

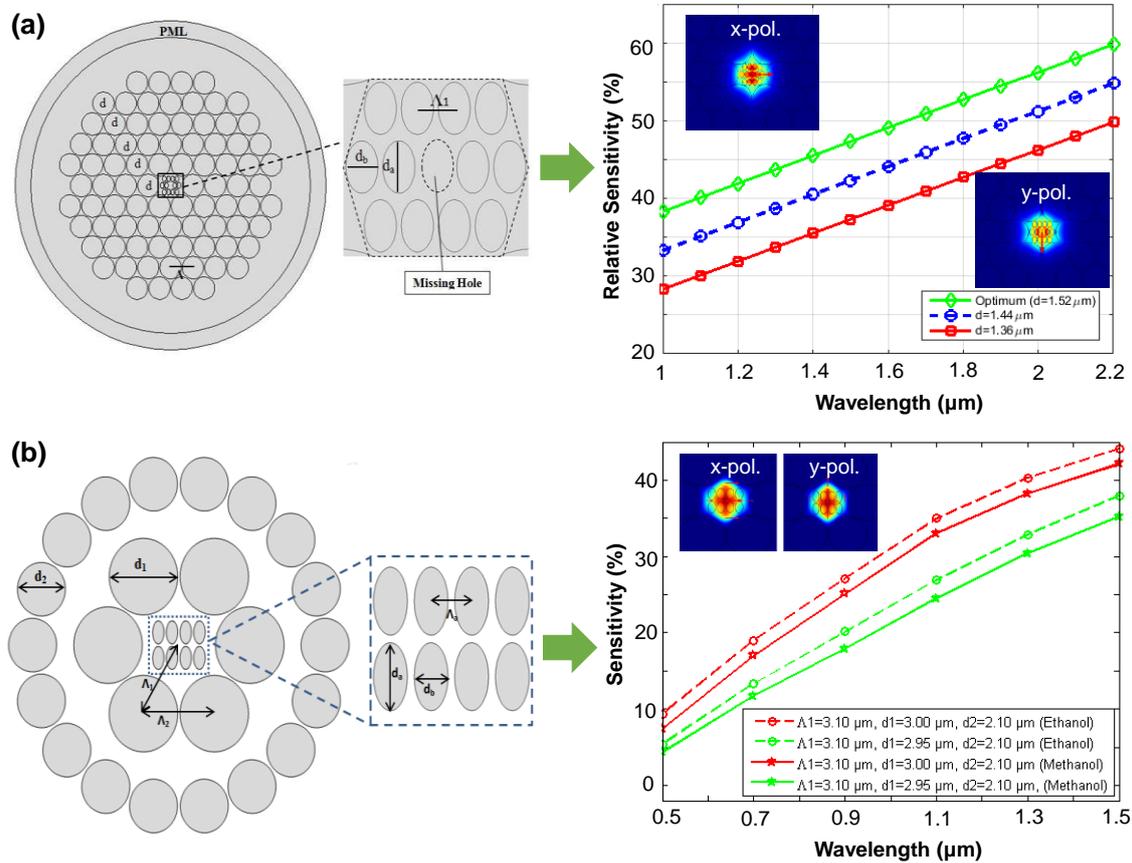

**Fig. 12.9** Analysis the diameter effect on relative sensitivity with (a) hexagonal PCF, (b) hybrid-PCF [12, 69].

### 12.4.3 *Air-filling Ratio Effects on Sensing*

Air filling ratio is the ratio between diameters of air-hole and pitch. The air filling ratio is another controlling parameter of relative sensitivity of PCF sensors. By changing both diameter and pitch at a certain ratio together air filling ratio changes. There exists a cut-off air-filling ration which is require to maintain. The cut off value of the air filling ratio is set as level of 0.95. Fig 12.10 illustrates that, relative sensitivity changes a lot due to small change of air filling ratio. The air filling ration can define by the following equation;

$$\text{Air filling ratio} = \frac{d \, (Diameter)}{\Lambda \, (pitch)} \qquad (12.12)$$

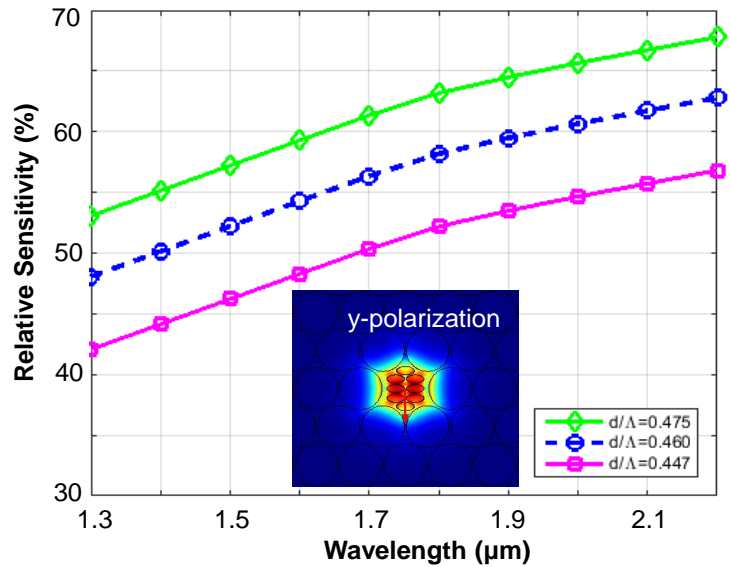

**Fig. 12.10** Analysis the effect of Air filling Ratio on relative sensitivity of PCF [47].

## 12.5  Core-Shape Effects on Sensing

In PCF based gas/ chemical sensor, core is a key parameter in terms of sensor performance. Generally, gases and chemicals are filled through the core as a result core shape has effects on sensing. According to Fig 12.11, it is visible that an elliptical hole in a rhombic orientation shows the higher sensitivity responses than the other orientation of air holes at the innermost core region.

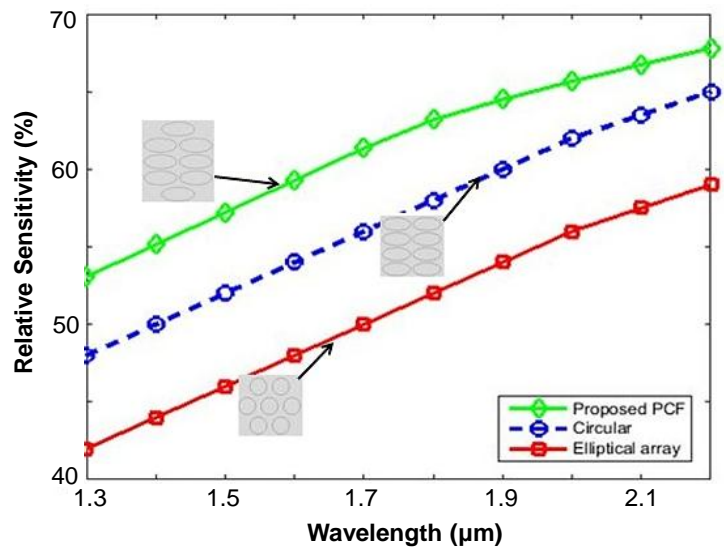

**Fig. 12.11** Core shape effect on Relative Sensitivity [47].

### *12.5.1  Hollow-Core PCF Based Sensors*

Recently, a simple hollow-core PCF (HC-PCF) where core is doped with different material has been reported for gas sensing [70] (see Fig. 12.12a). The operating wavelength varied from 0.8 - 1.60 μm to investigate the different guiding properties. At wavelength λ=1.60 µm, it shows the maximum sensitivity response of 19.94% (Fig. 12.12b) and at the same time it also reduces the confinement loss to $2.74\times10^{-4}$ dB/m.

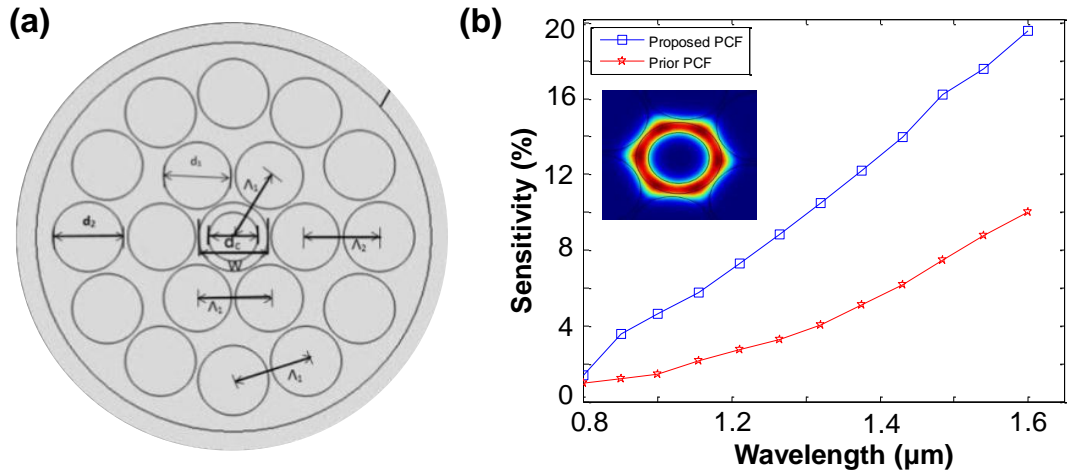

**Fig. 12.12** (a) Cross-section view of doped material based hollow-core PCF. (b) Sensitivity as a function of wavelength [70].

A hybrid structure photonic crystal fiber based gas sensor is presented in Fig. 12.13a, to detect toxic and colourless gases. Numerical study showed that sensitivity response of this hybrid-PCF sensor enhanced to 15.67% (Fig. 12.13b). The Confinement loss or Leakage loss decreased to $1.12\times10^{-7}$ dB/m by acquainting an octagonal ring of air holes in the outer cladding. This sensor works in wider range of wavelength from 0.8 to 2 µm.

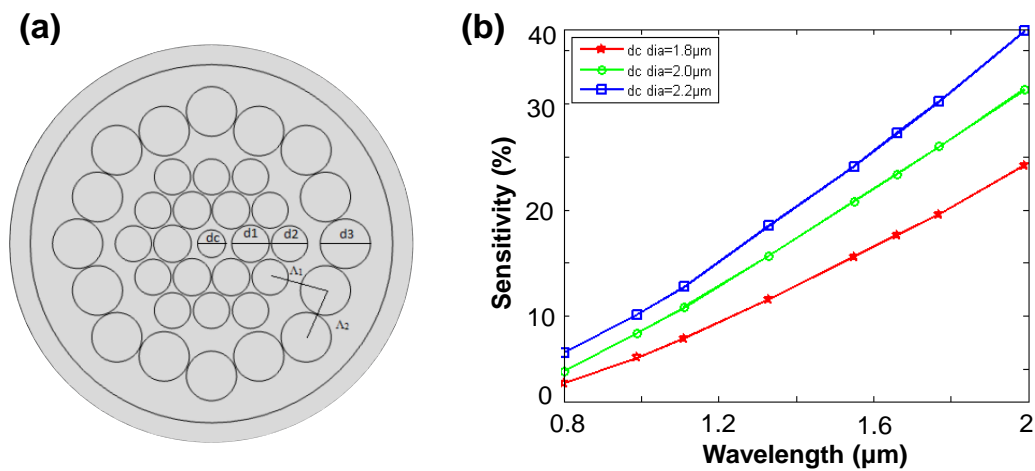

**Fig. 12.13** (a) Cross-section view of hollow-core based hybrid PCF. (b) Sensitivity as a function of wavelength.

Two different structures of HC-PCF have been presented in Fig. 12.14a & 14b [71]. The numerical result shows that hexagonal PCF (consist with six air-holes in the 1$^{st}$ ring) shows

2.22 times higher sensitivity responses compared to the octagonal PCF (consist with eight air-holes in the 1st ring) (Fig. 12.14c). Octagonal PCF also exhibits low confinement loss. These PCFs have been reported to sense the lower refractive index based gases (toxic/ flammable) at a wide range of wavelength 0.8 to 2 µm [71].

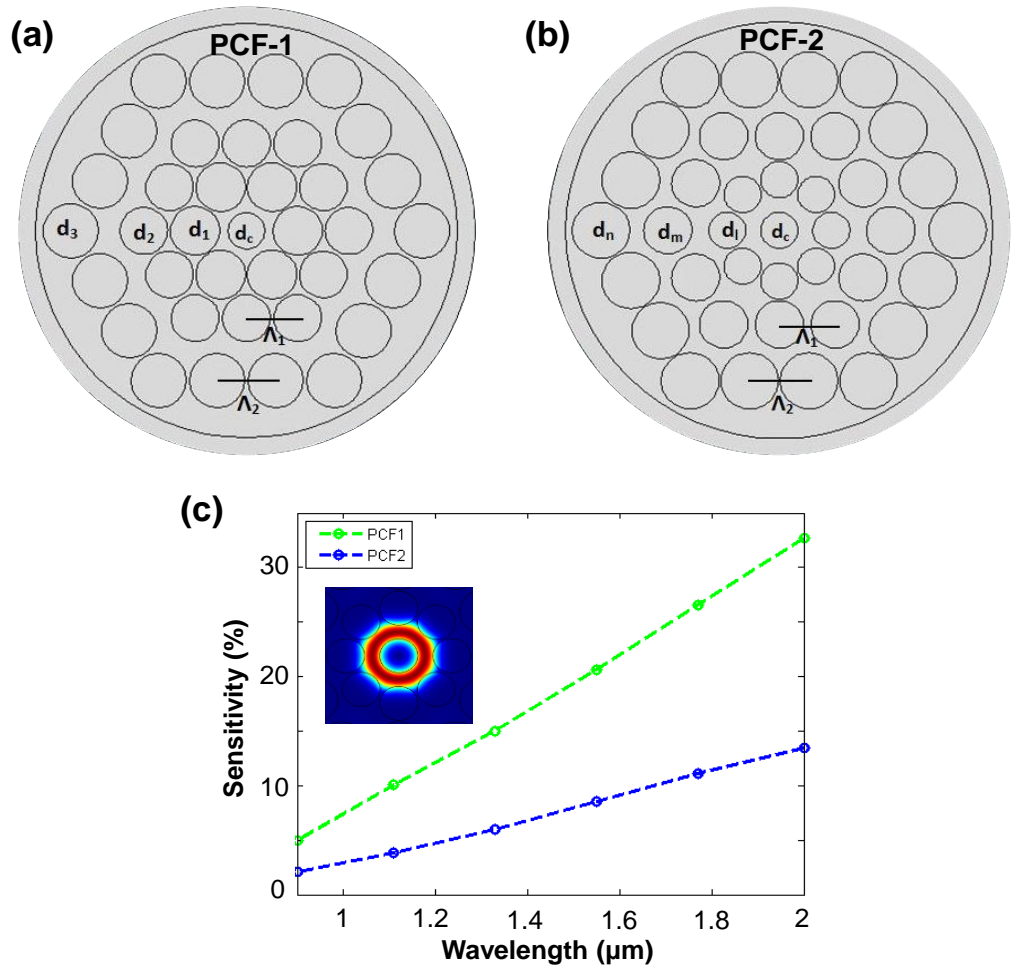

**Fig. 12.14** Cross-section view of (a) hexagonal, (b) octagonal PCF, and (c) comparison of two HC-PCFs based on sensitivity [71].

### *12.5.2  Slotted-Core PCF Based Sensors*

Slotted-core PCF widely used in Terahertz communication applications. However, recently it shows great interest in sensing applications as well. Recently, Asaduzzaman et al. reported a slotted-core PCF for gas sensing (Fig. 12.15a) [72]. Numerical result reveals that slotted-core PCF is more suitable for the sensing application and it shows the maximum relative sensitivity of 48.26%. By using slotted shaped air holes, relative sensitivity response increased a lot than the prior PCFs which is presented in Fig. 12.15b.

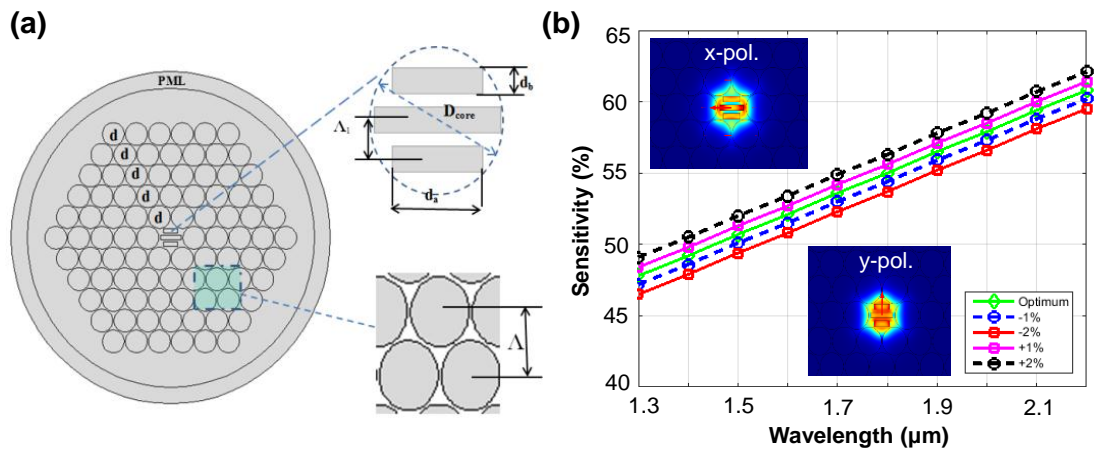

**Fig. 12.15** (a) Cross-section view of slotted-core PCF. (b) Sensitivity as a function of wavelength (inset shows the x & y-polarized mode) [72].

### 12.5.3 *Microstructured-Core PCF Based Sensors*

Recently, PCF with microarray pattern core has been reported by Asaduzzaman et al. [12] (Fig. 12.16a). The core region contains vertically arranged elliptical holes which enhanced the relative sensitivity responses. The elliptical holes are arranged in hexagonal shape with missing holes at centre. It shows the maximum sensitivity response of 43.7% (Fig. 12.16b). The structural geometric parameters also tuned to optimize the sensor performance.

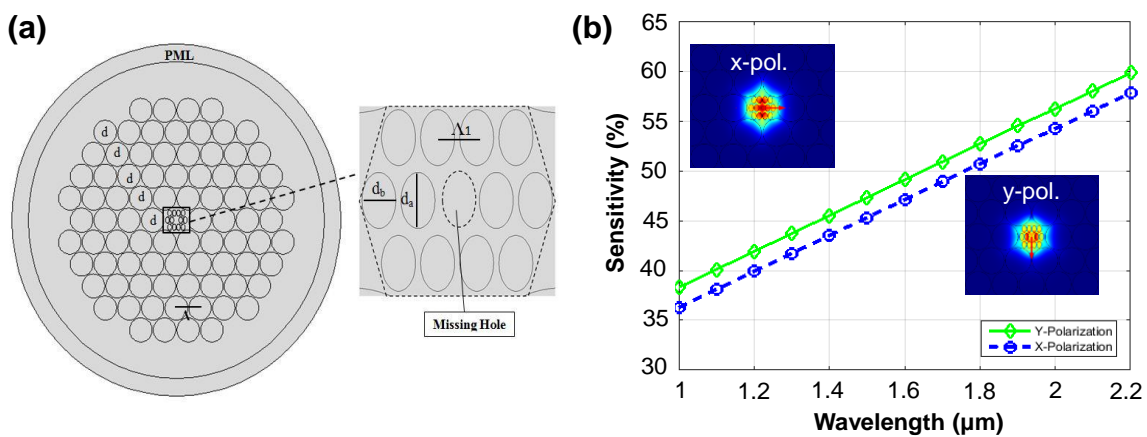

**Fig. 12.16** (a) Cross-section view of microarray core based PCF, and (b) sensitivity as a function of wavelength [29].

Asaduzzaman *et al.* [47] reported a micro-core PCF based gas sensor for detecting colourless or toxic gases and monitoring air pollution by measuring gas condensate components in production facilities (Fig. 12.17a). According to the computational results, the high relative sensitivity response of 53.07% is obtained at 1.33 μm wavelength for optimum parameters which is shown in Fig. 12.17b. Here, elliptical shaped holes are arranged in an elliptical manner which led to the higher sensitivity response than the previous reported sensor [47].

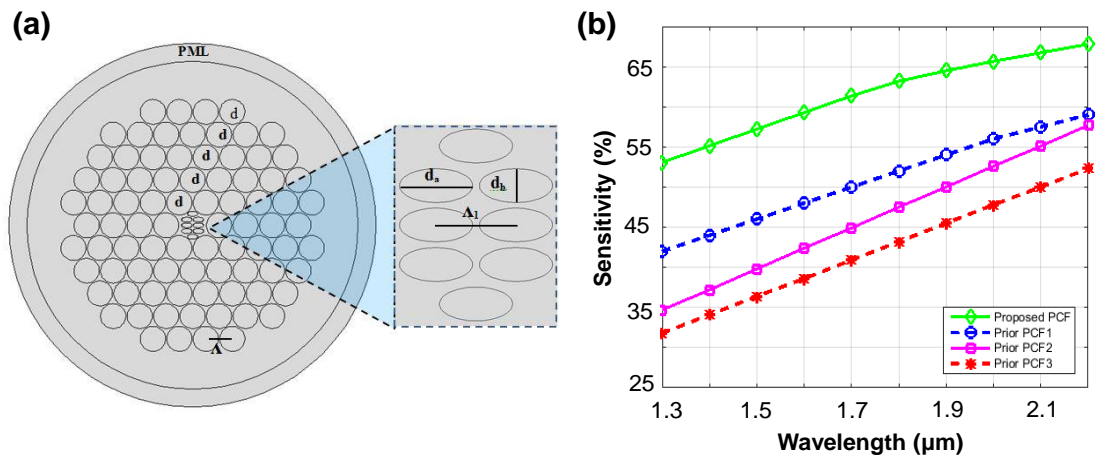

**Fig. 12.17** (a) Elliptical holes array core based PCF, and (b) sensitivity as a function of wavelength [47].

A circular photonic crystal fiber (C-PCF) based chemical sensor presented in Fig. 12.18a [73]. It investigated the detection of Ethanol and Propanol chemical compound. For both chemicals (n =1.354, n= 1.378), proposed C-PCF sensor shows higher relative sensitivity response compared to reported results in Fig. 12.12b and Fig. 12.14b (Fig. 12.18b).

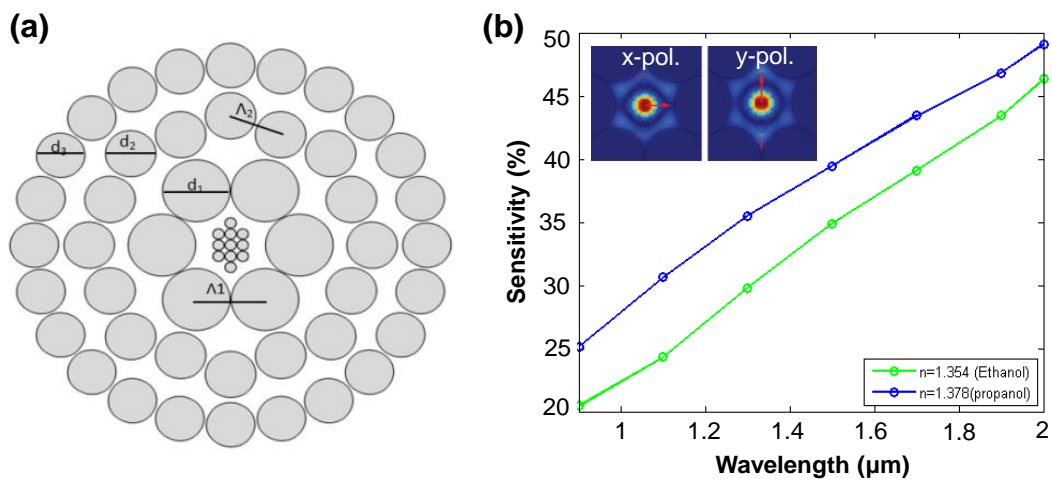

**Fig. 12.18** (a) Cross-section view of PCF with microarray-circular hole based core, and (b) sensitivity as a function of wavelength [73].

## 12.6  Cladding Effects on Sensing

Cladding is the outer layer of photonic crystal fibers which helps to concentrate light through the core region. Cladding has significant influence in reducing confinement loss and guiding light to pass through the center core which may results more power at core region. Shape of cladding means the different geometrical organization of air-holes surrounding the core region. Different cladding shape has very small impact on PCF based gas sensing. Different cladding structure based PCF gas sensor shows in Fig. 12.19. It is visible that hybrid shape PCF shows better relative sensitivity compared to the circular PCF [74]. Hybrid means asymmetric arrangement of air holes at cladding region such as, hexagonal, octagonal or in circular manner.

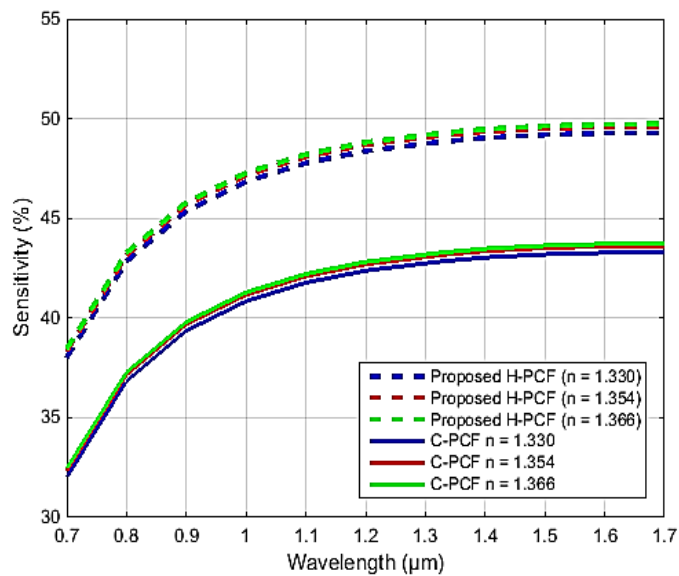

**Fig. 12.19** Cladding shape effects on relative sensitivity [74].

Cladding air filling ratio also affects the relative sensitivity. Fig. 12.20 reveals that higher air filling ratio introduces the higher relative sensitivity which increases with respect to wavelength.

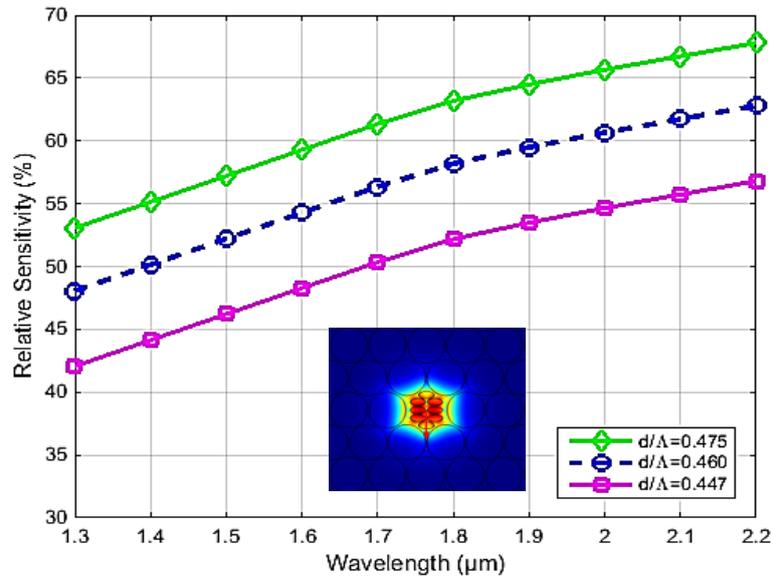

**Fig. 12.20** Cladding air filling ratio effect on relative sensitivity [47].

Cladding air-hole diameters are also responsible for changing in relative sensitivity. The larger air hole at outer layer of cladding reduces the confinement loss but, has very less effects on relative sensitivity. On the other hand, larger air-hole at inner most layer of cladding increases the sensitivity but, no significant effects on confinement loss. Besides, the increment of air-holes diameter in cladding increases the relative sensitivity which shows in Fig 12.21.

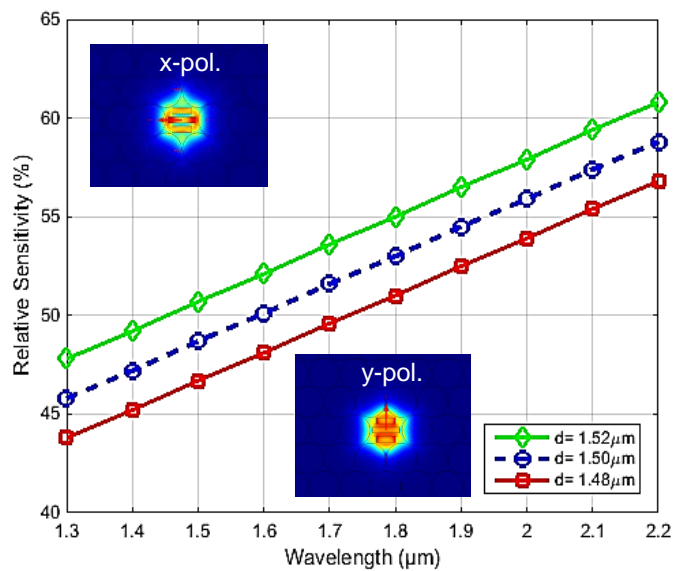

**Fig. 12.21** Analysis the cladding diameter effect on relative sensitivity [72].

Fig 12.22 exhibits the effects of inner layer and outer layer air holes effects on the relative sensitivity. According to Fig. 12.22b, it is clearly visible that first cladding layer air-holes diameter has significant effects on sensing. With the increase of first air-holes diameter, sensitivity increases significantly. On the contrary, third layer/ ring diameter variations has no significant effects on sensing (Fig. 12.22c).

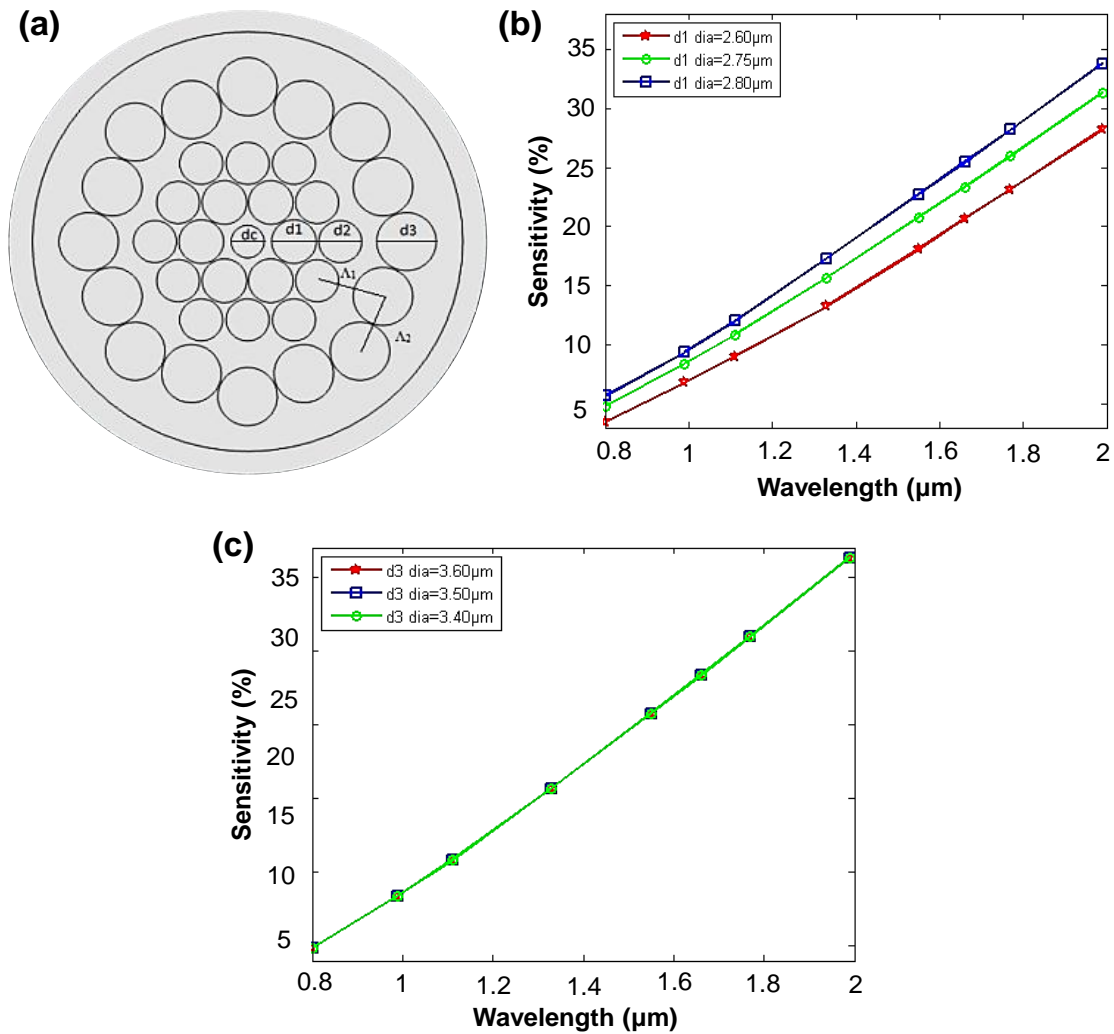

**Fig. 12.22** (a) Cross-section view of hybrid-PCF, (b) & (c) cladding diameter (layer based) effects on relative sensitivity.

According to Fig. 12.23, it is also agree with the previous argument that $1^{st}$ cladding layer is the important layer which has substantial effects on sensing performance. Fig. 12.23a is a two-ring based hybrid PCF. Fig. 12.23b shows the two different cases, one is for ethanol and another one is methanol gas. Both cases shows the same result that, due to increase of $1^{st}$ layer air-holes diameter sensitivity increases significantly.

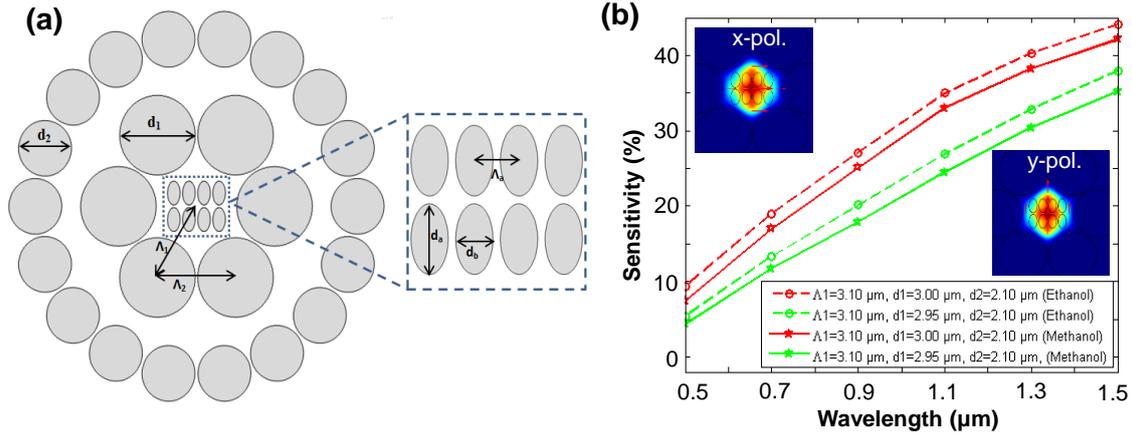

**Fig. 12.23** (a) Cross-section view of two ring hybrid PCF, and (b) sensitivity effects due to change of 1st layer cladding diameters [69].

## 12.7 Perfectly Matched Layer (PML) Effects on Sensing

Perfectly matched layer is an artificial layer that was primarily developed by the researcher Berenger [75]. It is the outermost layer of the PCF that enclosed the cladding region. Absorption boundary layer is needed to diminish the incident unwanted electromagnetic radiation which employ the role of absorption boundary condition (ABC) [76]. Generally, PCF based gas/ chemical sensors depth is set as 10% of the cladding region. The width of the PML doesn't have much effects on sensitivity response. However, it has effects on confinement loss measurement [77].

## 12.8 Fiber Background Material Effects on Sensing

Recently, Kawsar *et al.* [78], investigated the effect of background materials on sensing performance (Fig. 12.24). Fig. 12.24a shows the cross-section view of the reported sensor structure. It focuses the effect of background material on relative sensitivity at the operating wavelength 1.2 to 2 μm wavelength. The relative sensitivity rapidly increases with respect to wavelength. Normally, silica is used as a background material. However, they explored three different materials such as, Crown glass, Silica and Zblan as a background materials with tiny air holes in core regions. From Fig. 12.24b, it can be illustrated that although, initially, silica shows the highest sensing performance however, after crossing the operating wavelength 1.6 μm, zblan shows the maximum sensitivity. The core-guided fundamental mode with x and y-polarization for background material crown glass (i & ii), silica (iii & iv), and Zblan (v & vi) shown is Fig. 12.24c. Finally, it can be said that background materials also has strong influence on sensor performance.

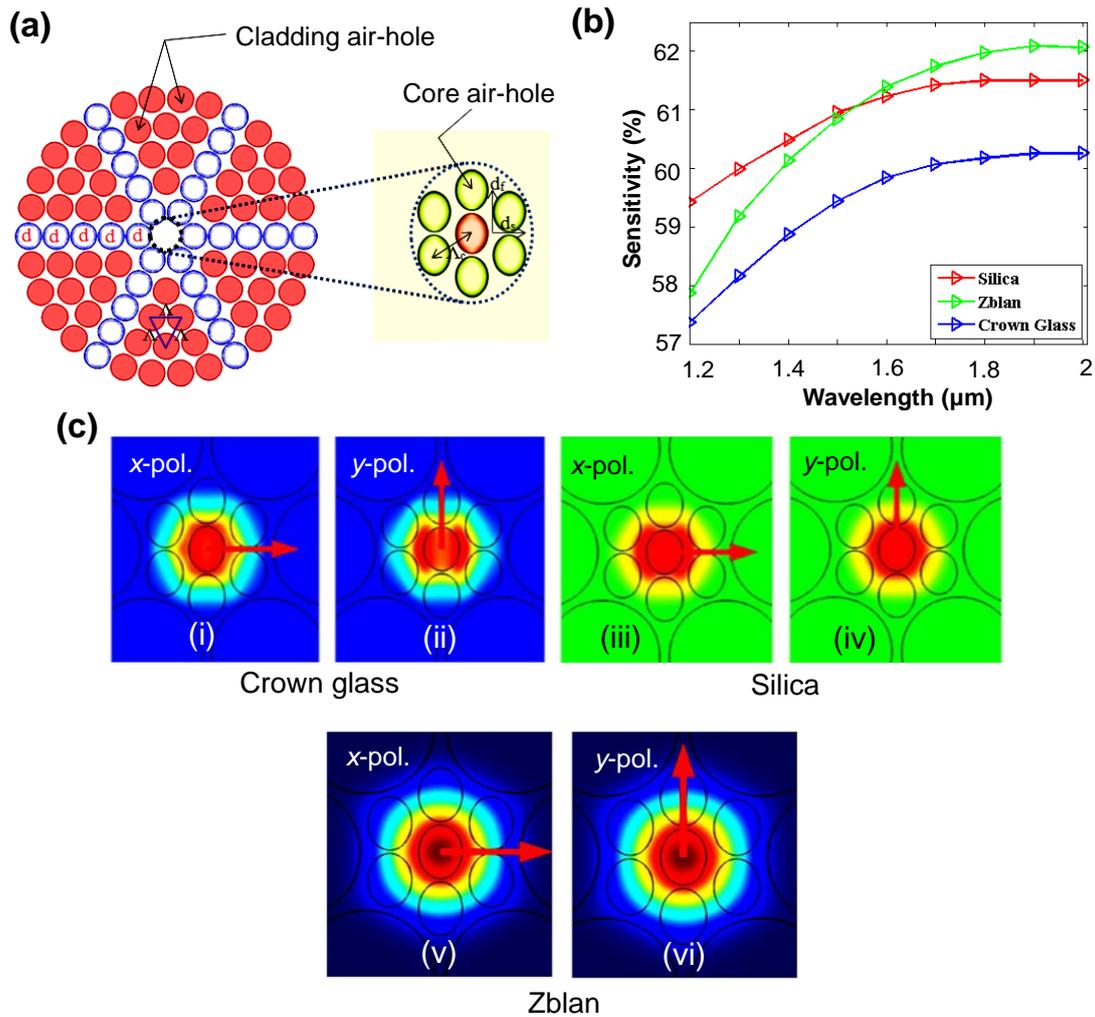

**Fig. 12.24** (a) Cross-section view of the PCF sensor, (b) Sensitivity as a function of wavelength, and (c) Core-guided fundamental mode with different background materials (i & ii) Crown glass, (iii & iv) silica, and (v & vi) Zblan [78].

## 12.9 Future Directions and Conclusions

In this chapter, we discuss the optical properties of PCF and working principle of PCF based gas/chemical sensors. We also extensively discuss the different types of PCF based gas/chemical sensors and also the effects of core-cladding and background materials on the sensing performance. The conventional PCF can be fabricated by following the standard stack-and-draw fiber fabrication method [79]. However, irregular PCF such as octagonal, decagonal, circular, kagome and hybrid structures fabrication are still challenging. Although the fabrication of irregular-PCFs is complex but nowadays it can be done by different fiber fabrication techniques. Sol-gel technique [80] is a modern innovation for PCF fabrication which enables to contrast diameter and pitch of any size of air holes. Besides, sol-gel casting

[81], extrusion [82], drilling [83] methods are suitable alternative to fabricate such irregular-PCF structure. The use of chemical and gas sensors in industries are becoming more popular. Moreover, toxic and harmful gases and chemicals are injurious and can be a cause of exploitation. Optical sensor more specifically photonic crystal fiber based sensors has proved its ability to detect toxic gases and chemicals. Although numerous computational works have been done for PCF based gas/ chemical sensing but, only limited number of works have been explored experimentally. Additional experimental investigations are required to practically implement the PCF based gas/ chemical sensor.

The future aspects of the PCF based gas/ chemical sensors are as follows-

- Need to simplify the PCF structure, so that it can easily be fabricated.
- For PCF based gas/chemical sensor, core region is very important because generally, gases and chemicals are flow through the core. As a result, need to make a suitable core-structure to absorb the light as much as possible.
- The propagation loss need to control which is important for the practical realization. Otherwise, light will immediately vanish after launching the light one end of the fiber as a result, it will not able to generate the measurable signal at the output end.
- Selective gas/ chemical infiltration is required in most of the reported PCF based gas/ chemical sensors. This is another challenge issue for practical realization. Its alternative solution could be external gas/ chemical sensing approach.